# Mirror Neuron; A Beautiful Unnecessary Concept

*J. N. Schad, PhD

[1]Retired LBNL (UCB) Scientist

The mirror neuron theory that has enjoyed continued validations till present was developed with no particular attention to the phenomenon of the vision itself that underlies it; perhaps it was thought of as a matter of course; understandably the perception of vision has always been thought to happen, naturally, as that for any of the other four senses. However, the reality that underlies this presumption of uniformity is by no means obvious; vision perception is based on remote sensing of the ecology, fundamentally different form that of the other senses, which have tactile stimulation origin (contact with matter). While its reality, as explicated as part of this work, explains why the above presumption is true, it also bears heavily on the mirror neuron theory: *the revelation of the nature of vision makes the prudently conceived of mirror neurons unnecessary*. Prima facie the extensive cognitive neurosciences investigation of primates and humans, over the past three decades, have experimentally validated the theory of mirror neurons which had been put forward early in the period (1980s and 1990s) [1, 2], based on the results of cognitive research experiments on the macaque monkeys. The concept was initially prompted by the fact that the brain activity patterns of the subjects were nearly similar, whether the activity was performed or observed by them. And presently, learning of various natures and empathy, and perhaps some aspects of survival, are ascribed to the operations of this class of additional neurons. Obviously the added complexity on the already complex field of neurosciences cannot be underestimated; and of course there are opponents of the theory and some profound questions have been raised [3]. Present work, though also in opposition is based on completely different ground: in this work I reason that all the results of the ingenious and grand efforts of the proponents of the theory can, not only find explanation in the context of the new *theory of vision* [4], but also provide further support for it. This new take of the phenomenon of vision is developed based on 1) the nature of the experimental methods that have succeeded in developing some measure of vision for the blinds, and 2) the inferences from the very likely nature of the computational strategy of the brain. The initial methods for vision restoration called *tactile vision substitution systems* (TVSS) [5, 6, 7], use stimulations of skin or other densely enervated areas, like tongue [8, 9], to create some degree of visual sensation. These and other experimental evidences [10], provide solid ground for the fundamental fact that vision perception, like touch, is a tactile phenomenon. And this revelation provides for sufficient explanations of the processes behind empathy, learning and perhaps other mental phenomena; and as such, the need for presumption of additional class of neurons is dispelled. Visual distal discernments, emotional and otherwise, are the final results of what in essence is the tactile stimulation of the eye receptor neurons by the *ecologically modulated* (both amplitude and frequency) light waves; or putting it equivalently, by the ecology affected photons (energy and intensity changed), which get treated (almost) as one's own sense of having touched all aspects of the ecological domain, in the realm one's mental states! The mental phenomena, which rendered the claim of the mirror neurons, are in essence the results of subjects beings variably *touched* by their ecology, through the *coherent tactile operation of all senses* (four already known as having tactile nature), all reconstructed on the substrate of a priory mental states.

Keywords: Mirror Neurons; Vision; Tactile; Electromagnetic Waves; Modulation; Encoding; Optogenetic

## INTRODUCTION

The idea of mirror neurons (the dubbing of the name was later in the research efforts), was promulgated by Dr. Giacomo Rizzolatti [1] and his colleague, of the university of Parma, Italy, as a result of their cognition experiments on the macaque monkeys (in 1980s and 1990s), which showed close brain signal patterns, mainly in the motor neuron, when an activity was either performed or observed by the subjects. Such works on primate and humans have been continuing and the article by Ferrari and Rizzolatti [2], --published in a special issue of the Philosophical Transaction following the workshop held in Ettore Mjorana (Sicily) on the occasion of 20$^{th}$ year of the discovery of the mirror neurons-- provides an account of the past works, and what path and progress that may lie ahead, in their views. Extent of the mirror neuron research, and the findings, that according to above investigators have "impacted disciplines outside neurosciences, such as psychology, ethology, sociology and philosophy, or that they had interested novelists and laymen," does not allow a deserving review of all the developments of the past 25 years research by the proponents and supports of the mirror neuron theory,-- many referenced in the *Phil. Transaction*-- and their few opponents [e.g., 3], especially, considering that they are not of direct relevance to the present work, which only draws from the widely known results of the associated experiments. These data, on the basis of which mirror neuron theory was postulated, find straight forward explanation in the light of a new vision theory [4], which was developed as a result of the early remedial efforts to help vision deprived or impaired individuals. These works that began with what was called tactile vision substitution system (TVSS) in the late 1960 [5, 6, 7, 8, 9], have continued onward; methods such as corneal [11], retinal and sub-retinal electrode implants [12], Optogenetic implants [13], and latest one, which drastically can improve the implant methods with the retina's neural code [14], have been developed. The latter method is based on an external stochastic retina simulator (a mathematical encoder formulated from animal retinal experimental data), which outputs (by means of a spectrum of blue light) into the ontogenetically activated ganglion cells, to create the train of visual spikes that is sent to the occipital lobe through the optic bundle. All these efforts have positive bearing on the vision theory. Also, the vision theory has ironically been partially validated by some *to the point* mirror neuron experimental works [15]. And actually, it also receives *unintentional partial conceptual* validation from the comprehensive mirror neuron theory validation experimental works using TMS (Trans-cranial Magnetic Simulation) [16], and fMRI results reported in Keysers et al. [15] (e.g.

Alaert, et al. [17]). Additional support may be drawn from the work of Aghajan et al. [18], in which experimental evidence for hippocampal placement activities for both distal visual cues and self-movements is reported. These various methods of visual system substitution that continue to be gaining improved measures of success, and (ironically) the mirror neuron experimental result, provide further conceptual framework for the validity of the (TVSS) vision theory which inherently falsifies the mirror neuron theory.

• Corresponding Author; email schadn5@Berkeley.edu

## The Synthesis

The mirror neuron theory that has enjoyed continued validations till present was developed with no particular attention to the phenomenon of the vision itself that underlies it; it was thought of as a matter of course; understandably the perception of vision has always been thought to happen, naturally, as that for any of the other four senses. However, the reality that underlies this true presumption of the uniformity is by no means obvious; vision perception is based on remote sensing of the ecology, fundamentally different form that of the other senses, which have tactile stimulation origin (contact with matter). While its reality, as explicated as part of this work, explains why the above presumption is true, it also bears heavily on the mirror neuron theory. To this end we need to explore beyond the knowledge of the biophysiological and optical aspects of the eyes; and the brain modalities where the trigger signals are processed: to know "how and where in the brain, vision perception occurs; pointedly in the context of knowledge of image developments in photography and the likes. Neurosciences' knowledge of the central nervous system, and the brain neurocomputational concepts, suggest that the brain neural processing (computational scheme) of eye-extracted afferent data (*somehow*) render vision perception. However, this does not provide any clue for the main puzzle; leaving the vision phenomenon vague in the least, for scientific minded, and mysterious to the most, as it has always been. It is not likely that further and more detailed understanding of the anatomy and physiology of the vision, and the related processes, would provide the answers. Efforts of decoding vision processes through brain imaging, though interesting [19], seemingly cannot provide any hint as to nature of the vision sensing.

We find the clue for the discovering of the nature of vision by analyzing how other four sense perceptions are realized in the beings: it starts from the afferent signals (action potentials) that develop from the physical contact of related nerves with matter; in solid, liquid or gas states forms. These signals,-- in the case of the touch (skin contact), generally being electrochemical conversion of the mechanical energy of vibrating (of varied frequencies) matter-- fire the brain neural net in order to make realization of the tactile and kinesthetic information (pressure, roughness, temperature, position force, and direction).The most important components of the ecology realization (perception, or figuring out) in *tactile* sensing, is the *simultaneity, intensity, and the correlation* (within the individual train of spike pattern of a single neuron and across those of the neighboring ones [20, 21]). The perception itself, how it comes about, is due to evolutionary schemes developed , *prior to the development of sight*, geared to species survival; *all based on direct (matter) contact with their ecology*. In such context, it is understood why congenital blinds would have sensation of their muscular-skeletal systems: it is realized from the train of action potentials, with the above characteristics, dispatched from all of their extremities through their somatic fibers. This is the way all the tactile *visual substitution systems* (TVSS) achieves their feat, though crudely; by subjecting a densely enervated body area to *frequent simultaneous and correlated pulsations* (electric, or piezoelectric), corresponding to the *varying ecological luminance* (varying photon energy intensity) that camera pixels relay through the digitally processed photoreceptor signals – generally manipulated using various schemes of integration or convolution. Recent studies [22] of brains segmental activities indicate *significant participation of visual cortex in the development of visual sensation in blinds,* while it is not the case for the sighted, for the activities of other senses.

For the sighted, a spectrum of varied luminance riches the eyes; that is, the returned light, with intensities from zero to almost100% of the visible part of electromagnetic waves, as a function of the ecology of the field of vision. The returned light can be thought of as *ecology modulated* light, in the likeness of other known modulation of electromagnetic waves, which are used to send *remote* information. Eyes' photoreceptors are *simultaneously* stimulated by this ecology information carrying light. And the corresponding energies of these Photons create time varying correlated (to various degrees) action potentials-- a physiologic *encoding* of the modulated waves; (coding) for the brain computations. The afferent signals thus created have captured the embedded ecologic information (edges, colors, expression of faces, actions, etc.), which is sent to the brain. The processing of this information engages the brain computational machinery [23, 24, 25, 26], mainly in the occipital lobe and less in other segments of the brain, to computes (to simulate) the ecology, from its (eye) stimulating luminance (the afferent signal content). The computed ecology is not the necessarily "its reality in itself," but it is what we have been habituated with (in the context of the a priori brain constructs), in the evolution process.

Considering the general consensus that computational machinery of the brain is networked based, its segmentation can be understood to be a function of complexity of the problems which variably engages it (in terms of the extent of the network), rather than necessarily the nature of the problem. This can be evinced in the versatility of both brain inspired scientific neural networks, and the digital computers which handle all ranges of problems, provided that the problem is well-posed and in the proper presentation (language encoding; training). *The plasticity of the brain also speaks to this fact*. Further, the action potentials being the trigger of for brain computations, its segmentation for the specific tasks must have been the results of its evolutionary energy efficiency schemes; and likely genetically sourced needed synaptic data (Patterns when needed) for various repeated computations (simulations). Therefore, it is not farfetched to imagine that the visual afferent signals, despite being massively data laden, would be *no different in nature* than other afferent signals from matter nerve stimulations of other senses; and be computationally processed, same as the *tactile signals,* by the brain, albeit through the deployment of more extensive circuitry (Occipital and other segments, engaging more than 20% of the network). Comparing the 2000 or so of the finger sensing nerves with more than one hundred million neurons of each eye, may explain why we do not see with our fingers. However, as known, the increased signaling (simultaneous intense and inherent correlation) that characterizes all tactile visual substitution systems, is the reason for their success in the development of some measure of vision sensation. Of course the brain neural engagements, despite some vision lobe participation engagement [10], is expectedly minuscule in TVSS, compared to what natural vision uses; and that is why only *very limited vision perception*, lacking details can be created. The retinal electrode implants considered *none-tactile*, have similar lack of visual definition due to the scarcity of the number of electrode pulsation points, while the better success of optogenic vision approach [12] is owed to the stimulation of massive variety of remnant and mostly non functioning neurons (photoreceptor, polar and ganglion). The ecology luminance encoding method of Nirenberg and Pandarinath [14], engages the occipital lobe more efficiently. Its advantage lies in that it replaces the limited electrodes of the implants with the millions of pixels in high resolution cameras, the output of which are almost directly fed into ganglion cells; a sizable number of healthy cells would be required for better simulation of the ecology. As a side note: if the encoding method overcomes the issues of the blue light, it is most likely that it will be optimally performing only in case of advance pigmentosas; presence of large numbers of remnant photoreceptors neurons may create encoding confusion, which may be difficult to correct.

Again, the limited success of the vision prosthetics, *even though they directly engage cortical neurons*, proves that all afferent signals, regardless of the source, are only distinguished in the “definition (degrees of detail)” of the perception they create, which is a function of the extent of the brain neural network that they engage--as mentioned before, brain segmentation must have been an evolutionary efficiency measure and needed synaptic grooming (training) that is needed for each sense, though each segment can provide computation assistance if any other segment is overwhelmed; technically in network size increase. Given such facts, one has to conclude that *vision sensation is the sensation of being touched by the environment and events, at hundreds of millions of nerve ending, rendering perception;* the familiar tactile perception of the ecology that the animate beings had developed in the pre- Cambrian era. This premise of tactile nature of vision allows for interpretations of all the results of the ingenious and grand efforts of the proponents and the promoters of the mirror theory, in terms of ecological touch perceptions; and thus preempts any need for the introduction of the new class of neurons. All complex mental phenomena, such as empathy, learning, tactile sympathy and blindsight, which necessitated the claim of the additional neurons are simply the results of subjects beings variably *touched* by what the impending ecology entails, through the *coherent tactile operation of all senses*, depending on the a priory mental states. And as such, the presumption of additional class of neurons is not justified in the context of the present knowledge of the brain functions.

The concept of the *seemingly metaphysical Mirror neurons* can be replaced with a *physical ones*, ones in which the neurons, *no different in essence from any other in the brain*, receive afferent signals resulting from eye neurons that are being essentially touched by the ecology, like those of from other senses, and computed likewise to create the needed perceptions; essentially a touch perception. What fMRI, and the likes, demonstrated is simply the fact that watching an event is very much like being physically involved in it, except for the lack of 1) the additional information that involvement of other senses would bring in, and 2) the additional engagement of many more brain checks and balances. Nonetheless, empathy, and antipathy, and most other distal emotional discernments, are the final results of tactile stimulation of vision sense by the *ecology* *modulated EM waves*, reflected form appearances, *which being afferent tactile signals*, get interpreted closely, as one's own, in the realm one's mental states!

## Conclusion

Given that the *modus operandi* of the brain which processes only tactile signals, for at least four of the five senses, to render their stimuli perceptions, and the limited vision created by the touch (TVSS), and other evidences, *it should not be strange that vision stimuli (the ecology) perception to be similarly a touch perception, though immensely detailed*: an evolution perfected sensing, akin to the familiar and evolution habituated of the other four senses. As such, the perception of the ecology happens *in the context of a unified tactile operation of all senses*. We are constantly touched (one way or the other) by the world and its events; and that the brain does not differentiate where the tactile signal are issued from; the difference is in the varied engagements of the brain structures and their functions which render various perceptions. Obviously combined participation of more senses engages more of the neurons of the brain, which are evinced in some of the fMRI brain activity observations. Finally, considering the natures of the phenomena of affection, language learning, songbirds learning to sing by listening, the blinds learning through audio-sensory, there should be little doubt about all perception to be based on tactile sensing process, regardless of the location of how, and from where the afferent tactile signals are relayed. The tactile basis of the ecology perception can account for all the experimental basis of the brilliantly thought of mirror neuron theory; thus dispelling any need for the addition of mirror neurons is dispelled.